\documentclass[preprint,onecolumn,superscriptaddress,nofootinbib]{revtex4-1}

\usepackage{amssymb}
\usepackage{enumerate}
\usepackage{amssymb}
\usepackage{amsmath}
\usepackage{feynmf}
\usepackage{slashed}
\usepackage{natbib}
\usepackage{graphicx}
\usepackage[pdftex,bookmarks,linktocpage,pdfpagelabels,plainpages=false,hyperfigures,linkcolor=blue,citecolor=blue]{hyperref} %for hypertext links
\hypersetup{colorlinks=true}

\usepackage{appendix}

\usepackage{mathrsfs,amssymb}  
\usepackage{cancel}
\usepackage[normalem]{ulem}

\newcommand\be{\begin{equation}}
\newcommand\ee{\end{equation}}

\newcommand\bea{\begin{eqnarray}}
\newcommand\eea{\end{eqnarray}}

\newcommand\mm{\mathcal{M}}

\begin{document}
\preprint{MI-TH-2032}
\bibliographystyle{apsrev4-1}

\title{Gamma Ray Signals from Cosmic Ray Scattering on Axion-Like Particles}

\author{James B.~Dent} 
\affiliation{Department of Physics, Sam Houston State University, Huntsville, TX 77341, USA}

\author{Bhaskar Dutta}
\affiliation{Mitchell Institute for Fundamental Physics and Astronomy,
   Department of Physics and Astronomy, Texas A\&M University, College Station, TX 77845, USA}

\author{Jayden L.~Newstead}
\affiliation{ARC Centre of Excellence for Dark Matter Particle Physics, School of Physics, The University of Melbourne, Victoria 3010, Australia}

\author{Alejandro Rodriguez}
\affiliation{Center for Neutrino Physics, Department of Physics, Virginia Tech University, Blacksburg, VA 24601, USA}
\affiliation{Department of Physics, The Pennsylvania State University, University Park, PA 16802, USA}

\author{Ian M. Shoemaker}
\affiliation{Center for Neutrino Physics, Department of Physics, Virginia Tech University, Blacksburg, VA 24601, USA}

\author{Zahra Tabrizi}
\affiliation{Center for Neutrino Physics, Department of Physics, Virginia Tech University, Blacksburg, VA 24601, USA}

\author{Natalia Tapia Arellano}
\affiliation{Center for Neutrino Physics, Department of Physics, Virginia Tech University, Blacksburg, VA 24601, USA}

\begin{abstract}

Dark Matter (DM) may be comprised of axion-like particles (ALPs) with couplings to photons and the standard model fermions. In this paper we study photon signals arising from cosmic ray~(CR) electron scattering on background ALPs. For a range of masses we find that these bounds can place competitive new constraints on the ALP-electron coupling, although in many models lifetime constraints may supersede these bounds.  In addition to current Fermi constraints, we also consider future e-Astrogram bounds which will have greater sensitivity to ALP-CR induced gamma-rays. 

%\com{Add discussion about perturbative couplings despite large cross sections...} \\

\end{abstract}

\maketitle

\section{Introduction}
The nature of the non-luminous dark matter (DM) is one of the greatest mysteries in physics. At present, everything that is known about DM has been derived on the basis of its gravitational interactions with ordinary matter. The race is on to detect any non-gravitational interactions of DM, which would be a significant step forward in our understanding of the most abundant type of matter in the Universe. 

A potential particle candidate for the DM is an axion-like particle (ALP), so called because they are light pseudoscalar particles which do not solve the strong CP problem of QCD. ALPs appear in some theories of physics beyond the standard model, e.g. string theory, as the pseudo Nambu-Goldstone boson resulting from the breaking of a $U(1)$ symmetry. This can induce a photon coupling through a mixing term of the form
\be
\mathcal{L} \supset -g_{a\gamma} a F_{\mu\nu}\Tilde{F}_{\mu\nu},
\ee
where {$F_{\mu\nu}$ is the electromagnetic field strength tensor with $\Tilde{F}_{\mu\nu}$ as its dual, $a$ is the ALP field and} $g_{a\gamma}$ is a coupling constant with units of ${\rm mass}^{-1}$.
In addition, ALPs may also couple to electrons via
\be
\mathcal{L} \supset - C_{ee}\frac{m_{e}}{\Lambda} a ~\bar{e} i \gamma_{5} e,
\ee
{with the dimensionless ALP electron coupling $C_{ee}$. }We will study the implications of both couplings in this paper.

A tremendous variety of experiments have been specifically designed to search for the
axion-photon coupling including helioscopes such as CAST \cite{Zioutas:1998cc,Anastassopoulos:2017ftl} and IAXO \cite{Irastorza:2013dav,Armengaud:2019uso}, haloscopes including Abracadabra \cite{Kahn:2016aff,Salemi:2019xgl}, ADMX \cite{Asztalos:2001tf,Du:2018uak}, CASPEr \cite{JacksonKimball:2017elr}, HAYSTAC \cite{Brubaker:2016ktl,Droster:2019fur}, light-shining-through-walls experiments including ALPSII \cite{Spector:2019ooq}, and interferometers \cite{Melissinos:2008vn,DeRocco:2018jwe} such as ADBC \cite{Liu:2018icu} and DANCE \cite{Obata:2018vvr}. Beam dump experiments can also manifest sensitivity to axion-photon couplings through axion decays or bremsstrahlung, including FASER \cite{Feng:2018noy}, LDMX \cite{Berlin:2018bsc,Akesson:2018vlm}, NA62 \cite{Volpe:2019nzt}, SeaQuest \cite{Berlin:2018pwi}, and SHiP \cite{Alekhin:2015byh}. Additionally, there are hybrid beam dump/haloscope designs such as PASSAT~\cite{Bonivento:2019sri}.
A summary of recent collider constraints can be found in~\cite{Bauer:2018uxu}. Neutrino and dark matter experiments such as
NOMAD~\cite{Astier:2000gx}, XMASS~\cite{Abe:2012ut}, EDELWEISS-III~\cite{Armengaud:2018cuy}, LUX, \cite{Akerib:2017uem}, PandaX-II~\cite{Fu:2017lfc}, Xenon1T~\cite{Aprile:2019xxb,Aprile:2020tmw}, and SuperCDMS~\cite{Aralis:2019nfa} have been leveraged as axion searches.  Geoscopes have also been proposed to study axion-electron couplings~\cite{Davoudiasl:2009fe}. The axion-photon coupling has been explored in direct detection experiments such as DAMA~\cite{Bernabei:2001ny}, EDELWEISS-II \cite{Armengaud:2013rta}, XMASS \cite{Oka:2017rnn}, and Xenon1T \cite{Aprile:2020tmw}. Axion-nucleon couplings are investigated from a solar axion flux produced through nuclear transitions, and can also be searched for through resonant absorption by laboratory nuclei~\cite{Moriyama:1995bz,Krcmar:1998xn,Krcmar:2001si,Derbin:2009jw,Gavrilyuk:2018jdi,Creswick:2018stb} including the CUORE Collaboration~\cite{Li:2015tsa,Li:2015tyq}, and the proposed GANDHI experiment~\cite{Benato:2018ijc}. The reactor experiment TEXONO has produced bounds on axion couplings~\cite{Chang:2006ug}, and upcoming neutrino experiments at nuclear reactor sites also having projected sensitivity in interesting axion mass and coupling parameter regions~\cite{Dent:2019ueq,AristizabalSierra:2020rom} .

Searches for ALPs can be direct, taking place in laboratories on Earth as listed above, or indirect, looking for the impact of ALPs on astrophysical objects. Traditional indirect detection searches of dark matter focus on its decay and/or annihilation products. However, the scattering of dark matter on electrons or nuclei is another possible probe~\cite{Hooper:2018bfw}. 

In this paper we perform an indirect search by using measurements from the Fermi Gamma-Ray Space Telescope~\cite{Ackermann:2014usa} and the Energetic Gamma Ray Experiment Telescope (EGRET) to constrain possible ALP interactions. We examine both electron and photon couplings to ALPs and find that the electron couplings produce bounds which are potentially competitive with bounds from other experiments. Additionally, we calculate future constraints from the planned E-Astrogram gamma-ray observatory.

\section{Calculational Framework}

In this paper we focus on axion-like particles (ALPs) as dark matter and examine the utility of cosmic ray-ALP scattering to probe the coupling of ALPs to ordinary matter.

\begin{figure}[t!]
\includegraphics[angle=0,width=.50\textwidth]{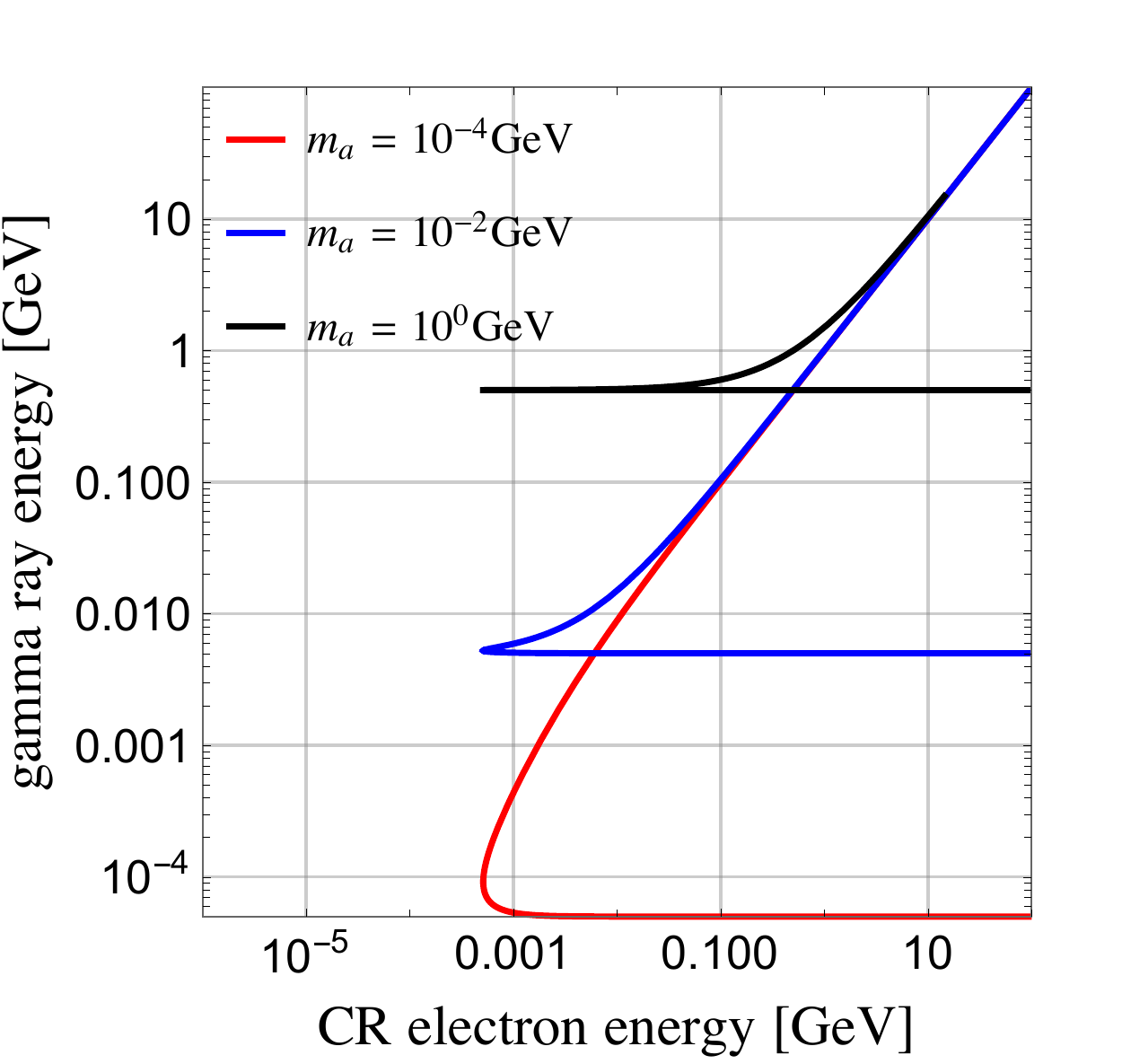}
\caption{Here we plot Eq.~\ref{eq:maxmin} in order to illustrate how the outgoing photon energy is influenced by the incoming CR energy and the ALP mass. At each CR energy the outgoing photon energy is constrained to lie to the right of the contour for fixed ALP mass. }
\label{fig:EgammaEcr}
\end{figure}

First notice that for a given incoming cosmic-ray energy, there is a minimum and maximum gamma-ray energy produced
\be
E_{\gamma}^{{\rm max/min}}= \frac{m_{a}^{2}+2m_{a} E_{e}}{2E_{e}+2m_{a}\mp 2 \sqrt{E_{e}^{2}-m_{e}^{2}}}\,,
\label{eq:maxmin}
\ee
where {$m_a$ is the ALP mass, $E_e$ is the electron energy, and} the maximum (minimum) occurs for forward (backward) scattering. We show in Fig.~\ref{fig:EgammaEcr} the dependence of the CR electron energy on the produced gamma-ray energy. Notice that at the highest energies, the gamma-ray energy is simply equal to the incoming CR electron energy. However for a fixed ALP mass, the gamma-ray energy will always be larger than half the ALP mass, $E_{\gamma} \ge m_{a}/2$. 

%%%%%%%%%%%%%
%%%%%%%%%%%%%

Having addressed the kinematics of ALP induced gamma-rays, we move on to the dynamics. We compute the expected gamma-ray flux from ALP-electron scattering as
\bea 
\frac{d \Phi_{\gamma}}{dE_{\gamma}} &=& \int_{V} dV \int_{E_{e}^{{\rm min}}} dE_{e} \frac{d^{2}  \Gamma_{e\rightarrow \gamma}}{dE_{e} E_{\gamma}}\\
&=& D_{{\rm eff}}~\frac{\rho_{DM}}{m_{a}} \int_{E_{e}^{{\rm min}}(E_{\gamma})} dE_{e} \frac{d \sigma_{e\gamma}}{dE_{\gamma}}~\frac{d \phi_{e}^{{\rm LIS}}}{dE_{e}}
\eea
where $\rho_{DM}\simeq 0.3$~GeVcm$^{-3}$ is the DM density, $\frac{d \phi_{e}^{{\rm LIS}}}{dE_{e}}$ is the electron flux of the local interstellar spectrum (LIS), and $D_{{\rm eff}}$ is the effective distance over which we include CRs as the source of ALP induced gamma-rays. For a sphere of radius 1 kpc centered on the Earth, and assuming an NFW profile, we take $D_{\mathrm{eff}}\simeq 1$ kpc~\cite{Bringmann:2018cvk}. Note that the lower bound on the integral, $E_{e}^{{\rm min}}(E_{\gamma})$ is the minimum electron energy to produce a given gamma-ray energy. This quantity can be found by inverting Eq.~\ref{eq:maxmin}. We use the LIS fluxes from Ref.~\cite{Boschini:2018zdv,Bisschoff:2019lne}, which are obtained from a fit to Voyager 1~\cite{Cummings_2016}, AMS-02~\cite{Aguilar:2014fea}, and PAMELA data~\cite{Adriani:2011xv}.

  We will examine the scattering process $a(k) + e(p) \rightarrow \gamma(q) + e(p')$ with only the pseudoscalar $g_{ae}$ coupling being non-zero. Note that it is possible to construct viable scenarios where the ALP-photon coupling can be minimized, see for example~\cite{Kaplan:1985dv,Craig:2018kne}. Performing the traces and computing all the scalar products, the squared amplitude for ALP-electron scattering is
\begin{widetext}
\bea
\label{eq:msquaredae}
\frac{1}{2}\sum_{spins}~|\mm|^2 &=&g_{ea}^2e^2\bigg[-\frac{2\left(m^4-m^2(2m_a^2+s+u)+su\right)}{(u-m^2)^2} 
\\\nonumber
&-& \frac{2\left(m^4-m^2(2m_a^2+s+u)+su\right)}{(s-m^2)^2}
\\\nonumber
&+&\frac{4\left(m^4+3m_a^2m^2-m^2(2s+t)+(s-m_a^2)(s+t)\right)}{(s-m^2)(u-m^2)}\bigg],
\eea
\end{widetext}
{where }the 1/2 factor arises from averaging over the initial electron spin (the axion has spin zero, thus only contributes a factor of one to the counting), and we have used the Mandelstam variable relation $s + t + u = 2m^2 + m_a^2$ to simplify the result. 

Next, in the case of nonzero photon-ALP coupling we find that the matrix element for $e + a \rightarrow e + \gamma$ is 
\bea
\label{eq:msquaredagamma}
\frac{1}{2}\sum_{spins}|\mathcal{M}|^2 &=& \frac{e^2g_{a\gamma\gamma}^2}{t^2}\left(m_a^2t(m^2+s) - m_a^4m^2 - t((s-m^2)^2 + st)-t(t-m_a^2)^2/2\right).
\eea
This matches with the results of Appendix B of~\cite{Yang:2012jra} when using the Mandelstam variable formulation.

With the matrix element squared in hand, in standard fashion, we then obtain the cross section for each process via
\bea
\frac{d\sigma}{dE_{\gamma}} = \frac{1}{32\pi m_a|\vec{p}|^2}\left(\frac{1}{2}\sum_{spins}|\mathcal{M}|^2\right)
\eea
where we have included the 1/2 factors as in Eqs.~(\ref{eq:msquaredae}) and (\ref{eq:msquaredagamma}). With these results in hand, we will turn to the observational constraints on these scattering processes.

%\com{TO DO: (1) XENONnT = 20 ton-yr, LZ = 5.6 ton * 1000 days, threshold =?. (2) Atomic coherence effect $\sim Z^{2}$, (3) Scalar interaction, dark photon to compare with [1709.07882], (4) BBN bounds weakened (https://arxiv.org/abs/1812.05605), LBECA (https://arxiv.org/pdf/2001.09311.pdf). }

\section{Results}
    \label{sec:results}
    
    We now confront the above predictions for ALP produced gamma-rays with experimental observations. Given that we need both large DM and CR densities, the galactic center region would be an obvious target. However, the strength of the ALP-CR interactions we are interested in are particularly large, and these may alter the ordinary expectation of cuspy density profiles. In addition, the signal-to-background ratio is presumably worse in this region given the large number of SM processes leading to photon production in this high-density region. 
    
    For the reasons argued above, we follow Ref.~\cite{Hooper:2018bfw} and utilize high-latitude data ($|b| > 20^{\circ})$ to constrain background ALPs. In this region the differences between cuspy and cored DM density profiles are less relevant, and the contributions from ordinary SM processes are suppressed. Observations of the Isotropic Gamma-Ray Background (IGRB) are a useful dataset in this regard. The IGRB is the residual component of the extragalactic background with diffuse galactic contributions and known point sources removed. This flux has been measured by the Fermi collaboration~\cite{Ackermann:2014usa} between (0.1 - 820) GeV. From statistical uncertainty alone, the total intensity attributed to the IGRB was found to be $(7.2 \pm 0.6) \times 10^{-6}$ cm$^{-2}$ s$^{-1}$  sr$^{-1}$. However, the collaboration estimates an additional $(15-30) \%$ systematic component driven by uncertainty in Galactic diffuse foregrounds. Lastly, we note that Ref.~\cite{DiMauro:2015tfa} has also used IGRB data to constrain DM, albeit in an annihilating DM framework.

%%%%%%%%%%%%%
\begin{figure}[t!]
\includegraphics[angle=0,width=.50\textwidth]{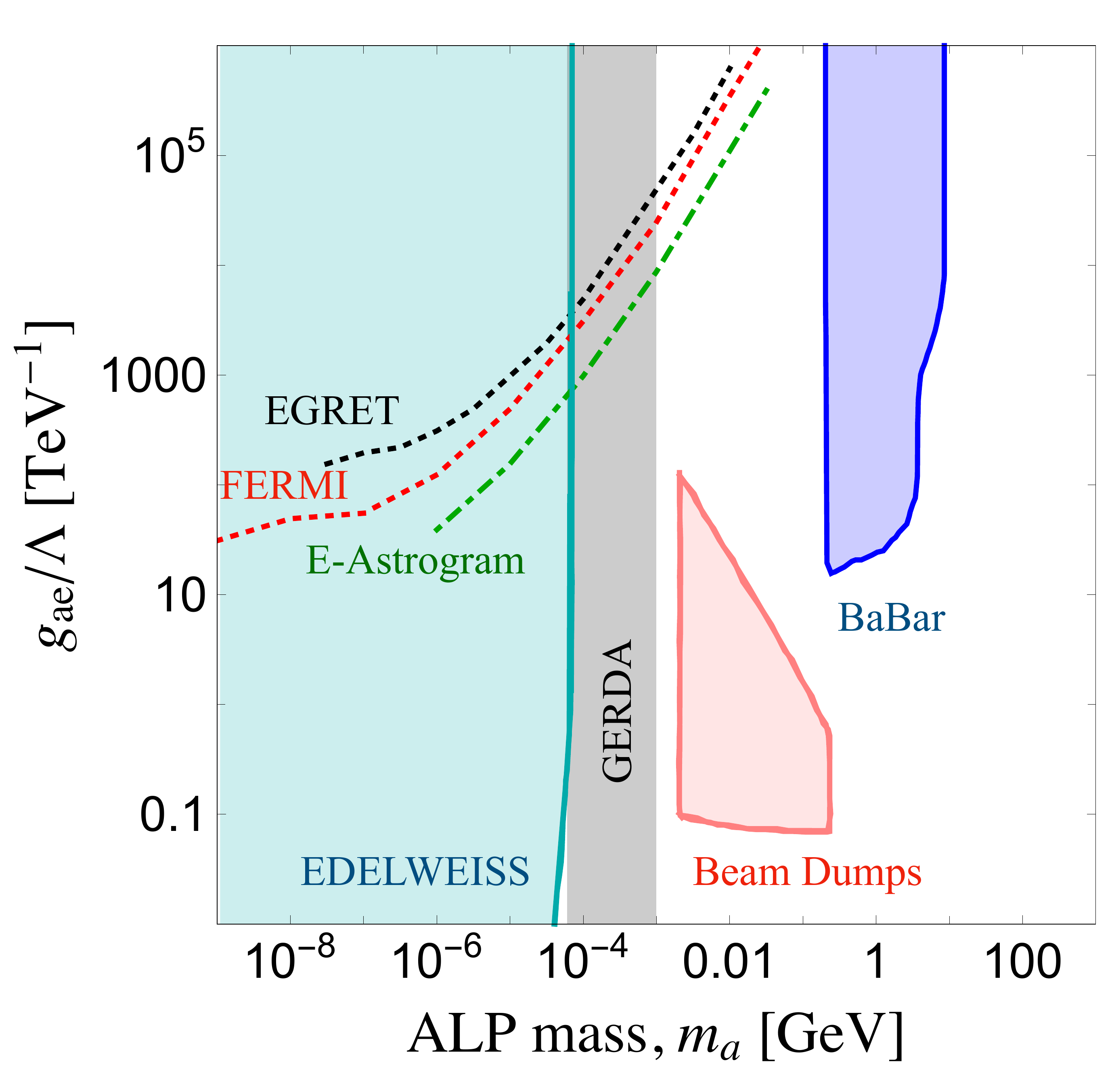}
\caption{Here we show our conservative constraints from Fermi data~\cite{Ackermann:2014usa} on the ALP-electron coupling, along with a variety of additional constraints from Babar~\cite{Lees:2014xha,Bauer:2017ris}, EDELWEISS-III~\cite{Armengaud:2018cuy}, and GERDA~\cite{GERDA:2020emj}, and beam dumps~\cite{Dobrich:2019dxc}. }
\label{fig:gaeplot}
\end{figure}
%%%%%%%%%%%%%    

%%%%%%%%%%%%%
\begin{figure}[t!]
\includegraphics[angle=0,width=.50\textwidth]{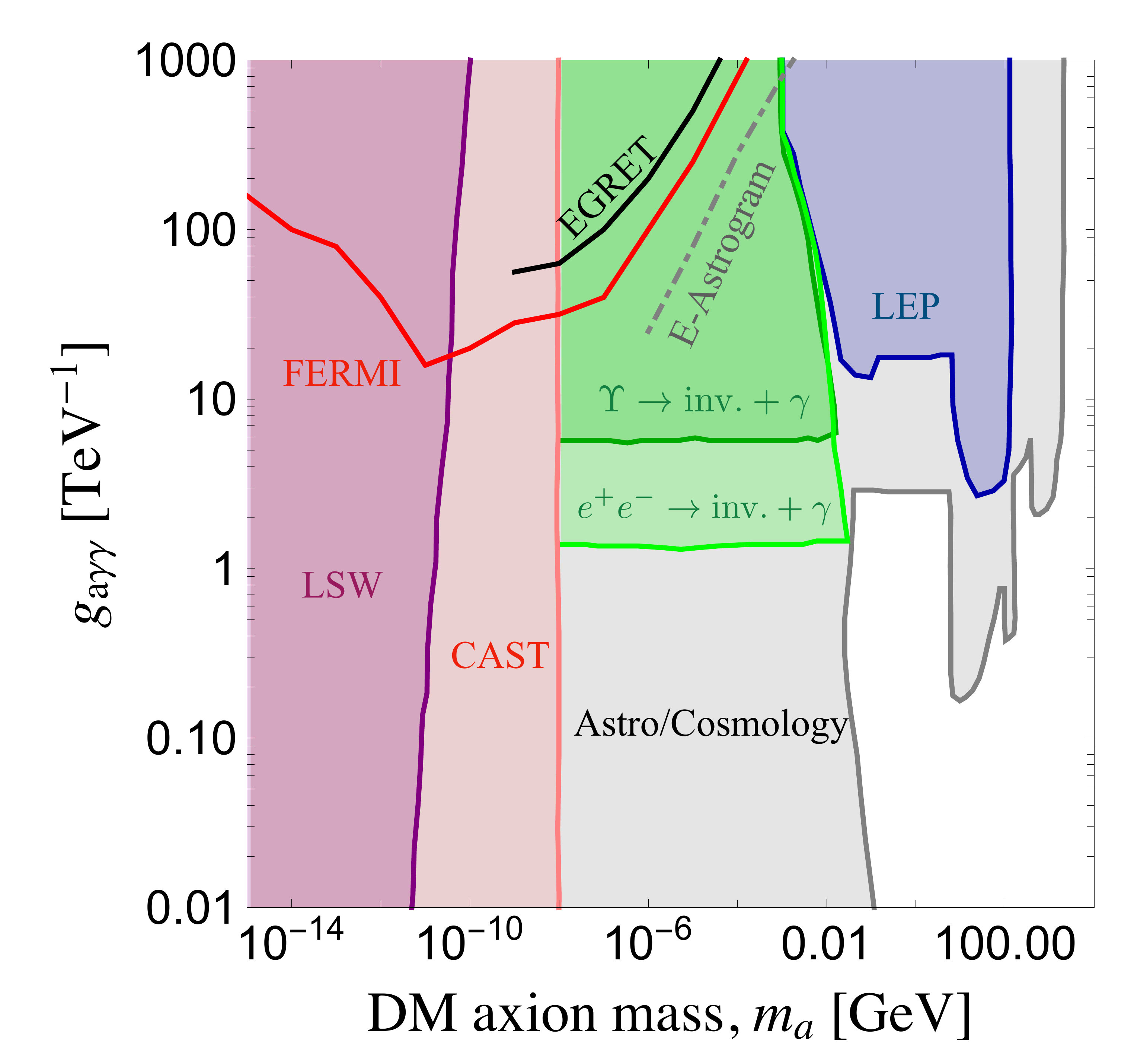}
\caption{Here we show the bounds on the axion-photon coupling from Fermi and EGRET, along with a projected sensitivity for E-Astrogram.}
\label{fig:Egamma}
\end{figure}
%%%%%%%%%%%%%

    To be conservative, we report bounds from Fermi by simply requiring that the predicted ALP induced gamma-ray flux not exceed the flux measured by Fermi. This is even more conservative than the approach adopted in Ref.~\cite{Hooper:2018bfw}. There it was argued that the new physics flux could at most comprise 20$\%$ of the IGRB since roughly 80$\%$ can be explained with known astrophysics involving non-blazar active galaxies~\cite{Hooper:2016gjy} and star-forming galaxies~\cite{Linden:2016fdd}.  
    
    Our results from the Fermi data~\cite{Ackermann:2014usa} are shown in Fig.~\ref{fig:gaeplot} and Fig.~\ref{fig:Egamma}. As can be seen, a region of $g_{ae}$ parameter space at relatively large ALP mass can be excluded which is not covered by existing experimental or observational data.  It should be noted, however, that this region has sufficiently large couplings that would naively lead to very rapidly decaying ALP DM, via $a\rightarrow \gamma \gamma $. Although the goal of the present work is not to construct a complete ALP model, it should be noted that various ``photophobic'' scenarios~\cite{Kaplan:1985dv,Agrawal:2017cmd} have been discussed in the literature which may alleviate such short lifetimes. We note that the bounds at low ALP mass can be improved by extrapolating the CR flux to higher energies. Thus with future high-energy CR data we expect the bounds at low ALP mass to improve considerably.
    
    Next one may wonder how much these constraints can be improved with future data. As a proof of principle, we look at the expected sensitivity of E-Astrogram~\cite{DeAngelis:2016slk}. E-Astrogram will dramatically improve on existing measurements in terms of statistics, but should still measure approximately the same central values of the photon flux in each bin already detected by other experiments. Thus in E-Astrogram's energy range 0.3 MeV to 3 GeV, we assume that the measured photon flux will have the same central values as already measured. Further, since this is an optimistic projection we will also assume that ordinary astrophysical explanations of known physics can be found for 90$\%$ of the measured flux, such that the ALP induced flux can at most account for 10$\%$ of the future IGRB flux seen by E-Astrogram. These projected constraints are shown in Fig.~\ref{fig:gaeplot} and Fig.~\ref{fig:Egamma}.  
    
    Lastly while we have focused on IGRB data to constrain ALP-CR interactions, one could potentially use other datasets. If for example a whole-sky search was conducted, this could enhance the signal by roughly a factor of $\sim 5$. Optimistically then, such a search may be able to improve over the bounds we have found here by a factor of $\sim 1/ \sqrt{5} \simeq 0.45$.

\section{Conclusions}
    \label{sec:conclusions}
    
We have investigated for the first time the production of photons from ALP-cosmic ray collisions. Through the use of FERMI gamma-ray observations of the Isotropic Gamma-Ray Background, we have found this process provides new competitive constraints on ALP-electron couplings, including probing new parameter space in the $g_{ae}-m_a$ plane. While here we have focused on the electron and photon couplings, cosmic rays could also be used to develop new constraints on the ALP-proton couplings. 

In addition to examining the bounds implied by present FERMI data on the IGRB, we have also made optimistic projections for the sensitivity of the future e-Astrogram observatory to ALP-induced gamma-rays. We have been conservative here in that we systematically underestimate the photon flux since we have only included diagrams with single outgoing photons. Secondly, we have only utilized existing data from Fermi, though one could exploit the predicted spatial morphology of the CR-ALP induced flux to impose some spatial cuts which could improve the signal-to-background ratio. Lastly we note that with future improvements in our understanding of the IGRB these bounds may be considerably improved.

\vspace{1cm}

{\bf \emph{Acknowledgements-  }} JBD acknowledges support from the National Science Foundation under Grant No. NSF PHY-1820801. The work of B.D. is supported in part by the DOE Grant No. DE-SC0010813.  The work of IMS, ZT, and NTA is supported by the U.S. Department of Energy under the award number DE-SC0020250./ JLN is supported by the Australian Research Council.

\bibliography{ref}

\end{document}